%% file: mgirardiprep1.tex
\documentclass[printer]{aa}
\usepackage{graphicx}
\usepackage{txfonts}
\def\lesssim{\mathrel{\hbox{\rlap{\hbox{\lower4pt\hbox{$\sim$}}}\hbox{$<$}}}}
\def\gtrsim{\mathrel{\hbox{\rlap{\hbox{\lower4pt\hbox{$\sim$}}}\hbox{$>$}}}}
\newcommand{\mincir}{\raise -2.truept\hbox{\rlap{\hbox{$\sim$}}\raise5.truept
\hbox{$<$}\ }}
\newcommand{\magcir}{\raise -2.truept\hbox{\rlap{\hbox{$\sim$}}\raise5.truept
\hbox{$>$}\ }}
\newcommand{\siml}{\raise -2.truept\hbox{\rlap{\hbox{$\sim$}}\raise5.truept
\hbox{$<$}\ }}
\newcommand{\simg}{\raise -2.truept\hbox{\rlap{\hbox{$\sim$}}\raise5.truept
\hbox{$>$}\ }}
\newcommand{\be}{\begin{equation}}
\newcommand{\ee}{\end{equation}}
\newcommand{\ba}{\begin{eqnarray}}
\newcommand{\ea}{\end{eqnarray}}
\newcommand {\h} {Mpc$\;$}

\newcommand {\hhh} {\;\mathrm{Mpc}}
\newcommand {\hh} {Mpc}
\newcommand {\ks} {km~s$^{-1} \;$}
\newcommand {\kss} {km~s$^{-1}$}
\newcommand {\m} {$M_{\odot} \;$}
\newcommand {\mm} {$M_{\odot}$}

\newcommand {\mqua} {$\times 10^{14}M_{\odot} \;$}
\newcommand {\mquaa} {$\times 10^{14}M_{\odot}$}

\newcommand{\degree}{\ensuremath{\mathrm{^\circ}}}
\newcommand{\arcm}{\ensuremath{\mathrm{^\prime}}}
\newcommand{\arcs}{\arcm\hskip -0.1em\arcm}
\newcommand{\dotarcs}{\,\rlap{\hbox{$\mathrm{^\prime\hskip-0.1em^\prime}$}}{\hbox{$.$}}\,}
\newcommand{\dotarcsss}{\,\rlap{\hbox{$\mathrm{^\prime\hskip-0.1em^\prime}$}}}
\newcommand{\dotsec}{\,\rlap{\hbox{$\mathrm{^s}$}}{\hbox{$.$}}\,}
\newcommand{\cluster}{RX\,J0152.7$-$1357$\,$}
\newcommand{\clusterr}{RX\,J0152.7$-$1357}
\begin{document}
   \title{Internal dynamics of the $z\sim0.8$ cluster \cluster
\thanks{Based in part on observations carried out at the European
Southern Observatory using the ESO Very Large Telescope on Cerro
Paranal (ESO programs 166.A-0701, 69.A-0683, and 72.A-0759) and the
ESO New Technology Telescope on Cerro La Silla (ESO program
61.A-0676).
}}

%
\author{M. Girardi\inst{1} 
\and R. Demarco\inst{2,3}
\and P. Rosati\inst{2}
\and S. Borgani\inst{1}
}
   \offprints{M. Girardi (girardi@ts.astro.it)}

\institute{
Dipartimento di Astronomia, Universit\`{a} degli Studi di Trieste, 
Via Tiepolo 11, I-34100 Trieste, Italy
\and  ESO-European Observatory, Karl-Schwarzschild-Str. 2, 
85748 Garching, Germany
\and Department of Physics \& Astronomy, 
Johns Hopkins University, 3400 N. Charles Street, Baltimore, MD 21218, USA
}
   \date{Received <date> /  accepted <date>}

\abstract{We present the results from the dynamical analysis of the
cluster of galaxies \clusterr, which shows a complex structure in its
X--ray emission, with two major clumps in the central region, and a
third clump in the Eastern region.  Our analysis is based on redshift
data for 187 galaxies.  We find that \cluster appears as a well
isolated peak in the redshift space at $z=0.836$, which includes 95
galaxies recognized as cluster members. We compute the line--of--sight
velocity dispersion of galaxies, $\sigma_{\rm V}=1322^{+74}_{-68}$
\kss, which is significantly larger than what {\bf is} expected in the
case of a relaxed cluster with an observed X--ray temperature of 5-6
keV.  We find evidence that this cluster is far from dynamical
equilibrium, as shown by the non Gaussianity of the velocity
distribution, the presence of a velocity gradient and significant
substructure.  Our analysis shows that the high value of $\sigma_{\rm
V}$ is due to the complex structure of \clusterr, i.e. to the presence
of three galaxy clumps of different mean velocity.  Using optical data
we detect a low--velocity clump (with $\sigma_{\rm V}=300$--500 \kss)
in the central South--West region and a high--velocity clump (with
$\sigma_{\rm V}$ $\sim 700$ \kss) in the Eastern region, well
corresponding to the South--West and East peaks detected in the X--ray
emission.  The central North--East X--ray peak is associated to the
main galaxy structure with a velocity which is intermediate between
those of the other two clumps and $\sigma_{\rm V}\sim 900$ \kss.  The
mass of the whole system within 2 \h is estimated to lie in the range
(1.2-2.2)$\times 10^{15}$\mm, depending on the model adopted to
describe the cluster dynamics. Such values are comparable to those of
very massive clusters at lower redshifts.  Analytic calculations based
on the two-body model indicate that the system is most likely bound,
currently undergoing merging.  In particular, we suggest that the
South--West clump is not a small group, but rather the dense
cluster--core of a massive cluster, likely destined to survive tidal
disruption during the merger.

}
\authorrunning{M. Girardi et al.} 
\titlerunning{Internal Dynamics of \cluster}
\maketitle
%

\section{Introduction}

Clusters of galaxies are visible tracers of the network of matter in
the Universe, marking the high-density regions where filaments of dark
matter join together. In the hierarchical scenario of large--scale
structure, clusters form via merging of smaller clumps and accretion
of material from large scale filaments (e.g., Borgani \& Guzzo
\cite{bor01}; Evrard \& Gioia \cite{evr02}). From the observational
side, signatures of past merging processes are found in cluster
substructure and evidences for ongoing cluster mergers  are rapidly
accumulating (e.g., B\"ohringer \& Schuecker \cite{boh02}; Buote
\cite{buo02}; Girardi \& Biviano \cite{gir02}; Evrard \cite{evr04}).

Over the last few years significant progress has been made to extend
the above studies from local to distant clusters.  Pioneering analyses
suggest that no evidence of dynamical evolution is shown by the
cluster population out to $z\sim 0.3$--0.4 (Adami et al. \cite{ada00};
Girardi \& Mezzetti \cite{gir01}; but cf. Plionis \cite{pli02}).  On
the other hand, $z>0.5$ clusters have more X--ray substructures than lower--$z$
clusters (Jeltema et al. \cite{jel05}) and most clusters identified at
$z\gtrsim 0.8$ show an elongated, clumpy, or possibly filamentary
structure (e.g., Donahue et al. \cite{don98}; Gioia et
al. \cite{gio99}; Rosati \cite{ros04}) thus suggesting that present
observations are approaching the epoch of cluster formation.  Our
results on \cluster at $z\sim 0.8$ add further insights on this issue.

The galaxy cluster \cluster was discovered in the ROSAT Deep Cluster
Survey (RDCS, Rosati et al. \cite{ros98}) in the ROSAT PSPC field
rp60000rn00 observed in January 1992. It was independently discovered
in the Wide Angle ROSAT Pointed Survey (WARPS, Ebeling et
al. \cite{ebe00}) and reported in the Bright SHARC survey (Romer et
al. \cite{rom00}). It appeared also in the list of X--ray extended sources
obtained from {\it Einstein} IPC data by Oppenheimer et
al. (\cite{opp97}).

The BeppoSax observations were used to derive a cluster X--ray
bolometric luminosity $L_{\rm{X,bol}}=(22\pm5)\times 10^{44}$
erg\,s$^{-1}$ ($h=0.5$ and $q_0=0.5$), and a gas temperature
$kT=6.46^{+1.74}_{-1.19}$ keV (Della Ceca et al. \cite{del00}).
\cluster is characterized by a complex morphology with at least two
cores, both in the optical and X--ray data as recovered by Keck
imaging and Beppo--SAX data (Della Ceca et al. \cite{del00}).
Observations with Chandra also show a complex structure in the
intra-cluster medium with the presence in the central cluster region
of two peaks in the X--ray emission $95\arcs$ apart (North-East:
R.A.=$1^{\mathrm{h}}52^{\mathrm{m}}44\dotsec18$,
Dec.=$-13\degree57\arcmin15\dotarcs84$; South-West:
R.A.=$1^{\mathrm{h}}52^{\mathrm{m}}39\dotsec89$,
Dec.=$-13\degree58\arcmin27\dotarcs48$ [J2000.0]), and a possible
third peak to the East
(R.A.=$1^{\mathrm{h}}52^{\mathrm{m}}52\dotsec42$,
Dec.=$-13\degr58\arcmin5\dotarcs52$ [J2000.0]), see Maughan et al.
(\cite{mau03}).  {\bf 
The existence of an Eastern peak was confirmed by spectroscopic VLT
data and an independent analysis of the Chandra data by Demarco et
al. (\cite{dem05}), who detect it at the $> 3\sigma$ c.l. in X-rays (see
their Fig.~1).}  Chandra
observations gave a gas temperature for the North-East and South-West
central X--ray clumps of $kT=5.5^{+0.9}_{-0.8}$ keV and
$kT=5.2^{+1.1}_{-0.9}$ keV, respectively {\bf (Maughan et al.
\cite{mau03})}.  A complex structure with several clumps is also shown
by the gravitational lensing analysis of Jee et al. (\cite{jee05}):
{\bf in particular, the mass clump A corresponds to the Eastern X--ray
peak.}

A number of evidences suggest that \cluster may be undergoing a
merger: the displacement between peaks of gas distribution and of
galaxy/dark matter distribution (Maughan et al.  \cite{mau03}; Jee et
al. \cite{jee05}); the possible presence of a shock front (Maughan et
al.  \cite{mau03}); the presence of galaxies showing a very recent
star formation episode (J\o rgensen et al. \cite{jor05}); the
segregation of star--forming and non star--forming galaxies probably
induced by the intra--cluster medium interaction (Homeier et
al. \cite{hom05}).

Demarco et al. (\cite{dem05}) have performed an extensive
spectroscopic survey of \cluster based on observations carried out with
FORS1 and FORS2 on the ESO Very Large Telescope, obtaining more than
200 redshifts in the cluster field. Their analysis shows that \cluster is
characterized by a large velocity dispersion, $\sim 1600$ \kss, and
indicates a very complex structure. In particular, the galaxy
populations inhabiting the regions around the three main X--ray peaks
are characterized by different kinematical behaviour, in agreement 
with a cluster merging scenario.

On the basis of Demarco et al.  data we further investigate the
internal dynamics of \clusterr.  The spatial and kinematical analysis
of member galaxies is a powerful way to detect and measure the
amount of substructure, to identify and analyze possible pre--merging
clumps or merger remnants (Girardi \& Biviano \cite{gir02} and
refs. therein).  This optical information is complementary to X--ray
information since galaxies and intra--cluster gas react on different
time scales during a merger (see, e.g., numerical simulations by
Roettiger et al. \cite{roe97}; Ricker \& Sarazin \cite{ric01};
Schindler \cite{sch02}).

The paper is organized as follows.  We describe member selection and
present our results for global properties of \cluster in Sect.~2. We
present our analysis of internal dynamics in Sect.~3. We discuss our
results suggesting a tentative picture of the dynamical status of
\cluster in Sect.~4. We summarize our results in Sect.~5.

Unless otherwise stated, we give errors at the 68\% confidence
level (hereafter c.l.)

Throughout the paper, we assume a flat cosmology with $\Omega_m=0.3$,
$\Omega_{\Lambda}=0.7$ and $H_0=70$ \ks Mpc$^{-1}$. For this
cosmological model 1 arcmin corresponds to $458$ kpc at the cluster redshift.

\section{Member selection and global properties}

Our data sample consists of the spectroscopic survey of \cluster
presented by Demarco et al. (\cite{dem05}), i.e.  187 galaxies with
available redshift (see their Tables~4 and 5). We assume a typical
redshift error of $8\times 10^{-4}$ according to {\bf the} authors
prescriptions.

The identification of cluster members proceeds in two steps, following
a procedure already used for nearby and medium--redshift clusters
(Fadda et al. \cite{fad96}; Girardi et al. \cite{gir96}; Girardi
\& Mezzetti \cite{gir01}).

\begin{figure}
\centering
\resizebox{\hsize}{!}{\includegraphics{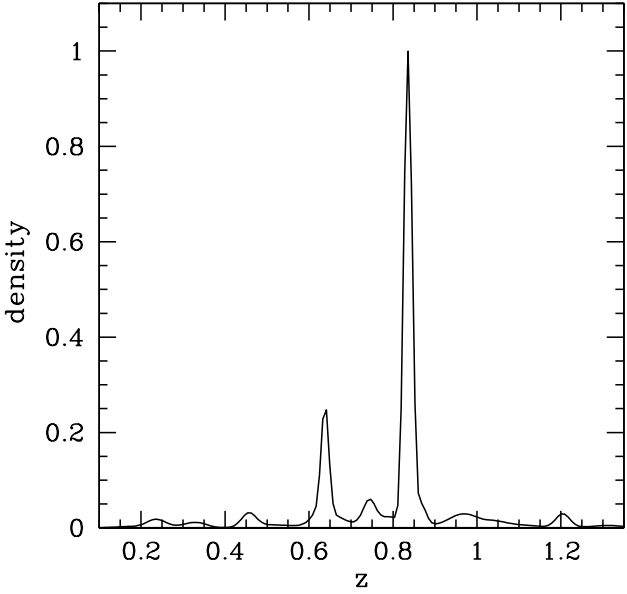}}
\caption
{The redshift galaxy density, as provided by the adaptive--kernel
reconstruction method. Unit on the y--axis 
is normalized to the density of the highest peak}
\label{figden}
\end{figure}

First, we perform the cluster--member selection in velocity space by
using only redshift information. We apply the adaptive kernel method
(Pisani \cite{pis93}) to find the significant ($>99\%$ c.l.)  peaks in
the velocity distribution.  This procedure detects \cluster as a well
isolated peak at $z=0.836$ assigning 103 galaxies considered as
candidate cluster members (see Fig.~\ref{figden}).  Out of non--member
galaxies, 61 and 23 are foreground and background galaxies,
respectively. In particular, a second significant peak of 31 galaxies
is shown at $z=0.638$ suggesting the presence of a foreground system.

All the galaxies assigned to the \cluster peak are analyzed in the
second step, which uses the combination of position and velocity
information.  We apply the procedure of the ``shifting gapper'' by
Fadda et al. (\cite{fad96}).  This procedure rejects galaxies that are
too far in velocity from the main body of galaxies and within a fixed
bin that shifts along the distance from the cluster center.  The
procedure is iterated until the number of cluster members converges to
a stable value.  We use a gap of $1000$ \ks -- in the cluster
rest-frame -- and a bin of 0.6 \hh, or large enough to include 15
galaxies.  As for the center we consider the position of the biweight
center, i.e. we perform the biweight mean-estimator (ROSTAT package; Beers et
al. \cite{bee90}) for ascension and declination separately: this
center is positioned between the North-East and South-West X--ray
peaks (see \S~1).  The choice of using either one of the two X--ray
peaks as cluster center does not affect the final results.

\begin{figure}
\centering 
\resizebox{\hsize}{!}{\includegraphics{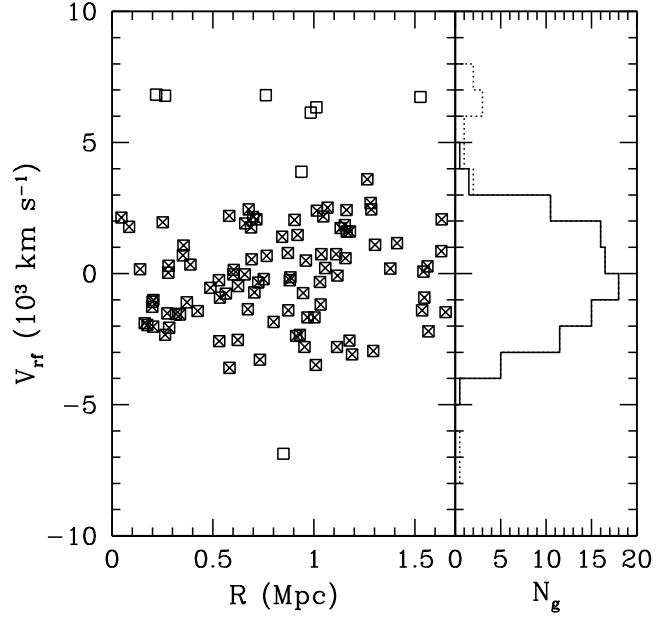}}
\caption
{Galaxies in the main peak of Fig.~\ref{figden}. Left panel: rest--frame
velocity vs. projected clustercentric distance; the 
application of the ``shifting gapper'' method rejects the galaxies
indicated by open squares.  Right panel: velocity distribution of all
and member galaxies (dotted and solid histograms, respectively). }
\label{figvd}
\end{figure}

The shifting--gapper procedure rejects eight galaxies to give 95
fiducial members.  The list of selected members corresponds to that in
Table~4 of Demarco et al. (\cite{dem05}), but excluding galaxies
$\#$306,509,557,650,895,1146,1239. Fig.~\ref{figvd} shows the plot of
rest-frame velocity $V_{\rm rf}=(cz-\left<cz\right>)/(1+\left<z\right>)$ vs. clustercentric
distance $R$ of galaxies in the main redshift peak.  Finally, we
recompute the biweight center on the 95 cluster members obtaining:
R.A.=$1^{\mathrm{h}}52^{\mathrm{m}}41\dotsec669$,
Dec.=$-13\degree\,57\arcmin\,58\dotarcs32$ (J2000.0).  Unless
otherwise stated, we adopt this as cluster center.

By applying the biweight estimator to cluster members (Beers et
al. \cite{bee90}), we compute a mean cluster redshift of
$\left<z\right>=0.8357\pm 0.0005$.  We estimate the line--of--sight (LOS)
velocity dispersion, $\sigma_{\rm V}$, by using the biweight estimator
and applying the cosmological correction and the standard correction
for velocity errors (Danese et al. \cite{dan80}).  We obtain
$\sigma_{\rm V}=1322^{+74}_{-68}$ \kss, where errors are estimated
through a bootstrap technique.

\begin{figure}
\centering 
\resizebox{\hsize}{!}{\includegraphics{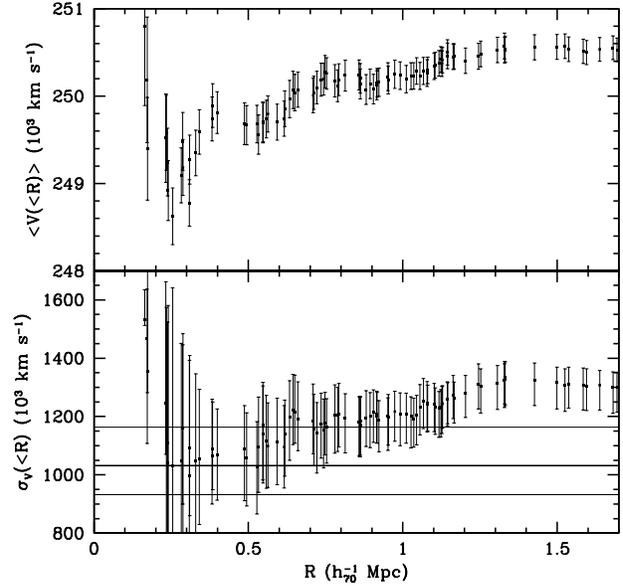}}
\caption
{Integrated mean velocity and LOS velocity--dispersion profiles (upper
and lower panel, respectively), where $\left<V\right>$ and $\sigma_{\rm V}$ at a given
(projected) radius from the cluster center is estimated by considering
all galaxies within that radius. The error bands at the $68\%$
c.l. are shown. In the lower panel, the horizontal line represent
X--ray temperature with the respective errors transformed in
$\sigma_{\rm V}$ imposing $\beta_{\rm spec}=1$ (see Sect.~4).}
\label{figprof}
\end{figure}

To evaluate the robustness of the $\sigma_{\rm V}$ estimate we analyze
the integral velocity dispersion profile (Fig.~\ref{figprof}).  The
value of $\sigma_{\rm V}(<R)$ sharply varies in the internal cluster
region.  A similar behaviour is shown by  the mean
velocity $\left<V(<R)\right>$ suggesting that a mix of clumps at different
redshifts is the likely cause for the high value of the velocity
dispersion rather than individual contaminating field--galaxies.  A
robust value of $\sigma_{\rm V}$ is reached in the external cluster
regions where the profile flattens, as found for most nearby clusters
(e.g., Fadda et al. \cite{fad96}).

The question of the presence of substructure is {\bf deferred} to the
following sections. Here we assume that the system is in dynamical
equilibrium to compute virial global quantities.

Following the prescriptions of Girardi \& Mezzetti (\cite{gir01}), we
assume for the radius of the quasi--virialized region R$_{\rm
vir}=0.17\times \sigma_{\rm V}/H(z) = 2.0$ \h (see their eq.~1 after
introducing the scaling with $H(z)$, see also eq.~ 8 of Carlberg et
al. \cite{car97} for $R_{200}$). Thus the cluster is sampled out to a
significant region, i.e.  $R_{\rm out}=0.82\times R_{\rm vir}$.

We compute the virial mass (Limber \& Mathews \cite{lim60}; see also,
e.g., Girardi et al. \cite{gir98})  using the data for the $N_g$ observed
galaxies:

\begin{equation}
M=3\pi/2 \cdot \sigma_{\rm V}^2 R_{\rm PV}/G-C ,
\end{equation}

\noindent where
$C$ is the surface term correction (The \& White
{\bf \cite{the86}}), and
$R_{\rm PV}$, equal to two times the (projected) harmonic radius, is: 

\begin{equation}
R_{\rm PV}=N_g(N_g-1)/(\Sigma_{i\ne j} R_{ij}^{-1}),
\end{equation}

\noindent 
where $R_{ij}$ is the projected distance between two galaxies.

The estimate of $\sigma_{\rm V}$ is generally robust when computed
within a large cluster region (see Fig.~\ref{figprof} for \cluster and
Fadda et al.  \cite{fad96} for other examples).  The value of $R_{\rm
PV}$ depends on the size of the sampled region and possibly on the
quality of the spatial sampling (e.g., whether the cluster is
uniformly sampled or not).  Here we obtain $R_{PV}=(1.45\pm0.05)$ \hh,
where the error is obtained via a jacknife procedure.  The value of
$C$ strongly depends on the radial component of the velocity dispersion
at the radius of the sampled region and could be obtained by
analyzing the velocity--dispersion profile, although this procedure
would require several hundreds of galaxies.  We apply the correction
obtained in the literature by combining data of many clusters sampled out
to about $R_{\rm vir}$ ($C/M_{\rm V}\sim20\%$, Carlberg et
al. \cite{car97}; Girardi et al. \cite{gir98}).  We obtain $M(<R_{\rm
out}=1.65 \hhh)=(2.2\pm0.3)\times 10^{15}$\mm.  Calling into question
the quality of the spatial sampling, one could use an alternative
estimate of $R_{\rm PV}$ on the basis of the knowledge of the galaxy
distribution. We assume a King--like distribution, with parameters
typical of nearby/medium--redshift clusters: a core radius
$R_C=1/20\times R_{\rm vir}$ and a slope--parameter $\beta_{\rm
fit}=0.8$, i.e. the volume galaxy density at large radii goes as
$r^{-3 \beta_{fit}}=r^{-2.4}$ (see G98 and Girardi \& Mezzetti
\cite{gir01}).  We obtain $R_{\rm PV}=1.25$ \hh, with a $25\%$ error,
thus in agreement with the above direct estimate. The mass is then
$M(<R_{\rm out})=(1.9\pm0.5)\times 10^{15}$\mm, in good agreement with
our first estimate.

We can use the second of the above approaches to obtain the mass
within the whole assumed virialized region, which is larger than that
sampled by observations, $M(<R_{\rm vir}=2.0 \hhh)=(2.2\pm0.6)\times
10^{15}$\mm.

\section{Dynamical analysis}

\subsection{Velocity distribution}

We analyze the velocity distribution to look for possible deviations
from Gaussianity that could provide important {\bf signatures} of complex
dynamics. For the following tests the null hypothesis is that the
velocity distribution is a single Gaussian.  We base our analysis on
 shape estimators, i.e. the kurtosis and the
skewness. As for the kurtosis, we find $K=2.04\pm0.49$, that indicates
a $\sim 2\sigma$ departure from a Gaussian distribution (reference
value $K=3$).  In addition, we compute the scaled tail index ($STI$),
which also measures the shape of a distribution, but is based on order
statistics of the dataset instead of its moments (see, e.g., Beers et
al. \cite{bee91}).  This estimator, $STI=0.860$, indicates that the
tails are underpopulated if the parent population is really a single
Gaussian with a c.l.  between $\sim 90\%$ and $\sim 95\%$, (see
Table~2 of Bird \& Beers \cite{bir93}). Finally, also the W--test
(Shapiro \& Wilk \cite{sha65}) rejects the null hypothesis of a
Gaussian parent distribution at the $98\%$ c.l..

\begin{figure}
\centering 
\resizebox{\hsize}{!}{\includegraphics{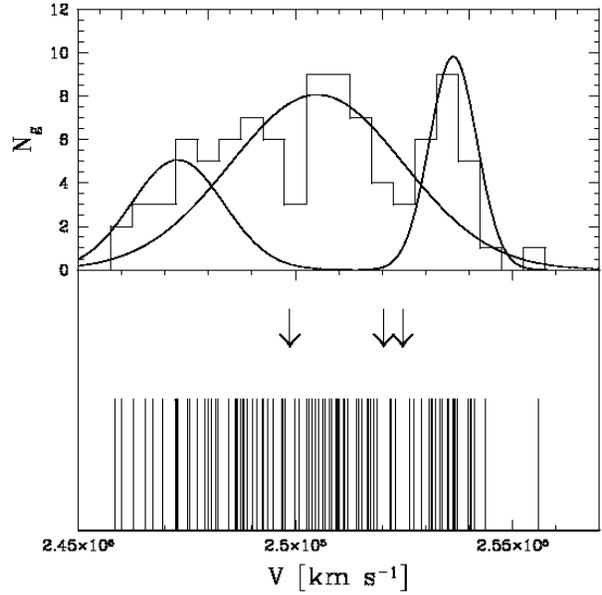}}
\caption
{{\bf Velocity distribution 
of radial velocities for the
95 cluster members. Bottom panel:} stripe density plot 
where arrows indicate the position of significant
gaps. The first gap lies between 249757 and 249997 \kss, the second
between 251886 and 252185 \kss, and the third between 252305 and
252635 \kss. {\bf Top panel: velocity histogram with a binning of 500 \ks
with superimposed the three Gaussians corresponding to KMM1a, KMM1b,
and
KMM2 in Table~\ref{tab1}.}}
\label{figstrip}
\end{figure}

Then we investigate the presence of gaps in the distribution, which
can be the signature of subclustering.  {\bf A weighted gap in the
space of the ordered velocities is defined as the difference between
two contiguous velocities, weighted by the location of these
velocities with respect to the middle of the data. We obtain values
for these gaps relative to their average size, precisely the midmean
of the weighted-gap distribution.}  We look for {\bf normalized} gaps
larger than 2.25 since in random draws of a Gaussian distribution they
arise at most in about $3\%$ of the cases, independent of the sample
size ({\bf Wainer and Schacht \cite{wai78}; see also} Beers et
al. \cite{bee91}). Three significant gaps (2.312, 2.366, 2.395) in the
ordered velocity dataset are detected (see Fig.~\ref{figstrip}).  From
low to high velocities the dataset is divided in parts containing 39,
29, 3, and 24 galaxies: thus the gaps individuate substantially three
main subsets.

In order to detect the presence of groups within our velocity dataset
we use the Kaye's mixture model (KMM) test (Ashman et
al. \cite{ash94}).  The KMM algorithm fits a user-specified number of
Gaussian distributions to a dataset and assesses the improvement of
that fit over a single Gaussian. In addition, it provides the
maximum-likelihood estimate of the unknown n-mode Gaussians and an
assignment of objects into groups. KMM is most appropriate in
situations where theoretical and/or empirical arguments indicate that
a Gaussian model is reasonable. This is valid in the case of cluster
velocity distributions, where gravitational interactions drive the
system toward a Gaussian distribution.  However, one of the major
uncertainties of this method is the optimal choice of the number of
groups for the partition. Moreover, only in mixture models with equal
covariance matrices for all components the algorithm converges, while
this is not always true for the heteroscedastic case (see Ashman et
al. 1994, for further details).

\begin{figure}
\centering 
\resizebox{\hsize}{!}{\includegraphics{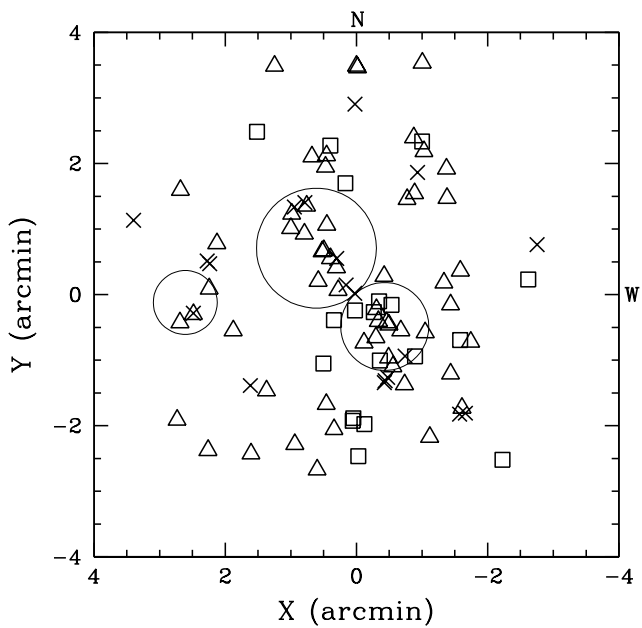}}
\caption
{Spatial distribution on the sky of the 95 member galaxies.  Open
symbols and crosses indicate galaxies assigned to KMM1 and KMM2
groups, respectively (see text). Squares and triangles indicate KMM1a
and KMM1b groups, respectively.  The plot is centered on the cluster
center defined in Sect.~2. Three circular regions, corresponding to
regions of extended X--ray emission are indicated, too (see Fig.~1 by
Demarco et al. \cite{dem05}).}
\label{figkmm}
\end{figure}

Our search for significant gaps suggests the presence of two Gaussians
(separated by the two very close second and third gaps at $\sim
252\times10^3$ \kss) or possibly three Gaussians (corresponding to the
three main subsets).  In the homoscedastic case the KMM algorithm fits
a two--group partition by rejecting the single Gaussian at the
$97.4\%$ c.l. (as obtained from the likelihood ratio test). The
three--group partition is fitted at the $97.9\%$ c.l.. In the
heteroscedastic case we use the results of the gap analysis to determine
the first guess and we fit two velocity groups around the guess
mean--velocities of $249\times10^3$ and $254\times10^3$ \kss. The
algorithm fits a two--group partition at the $99.4\%$ c.l. Similarly,
we fit three velocity groups around the guess mean--velocities of
$247\times10^3$, $250\times10^3$, and $254\times10^3$ \ks to obtain a
three--group partition at the $97.2\%$ c.l.  The high probability
value obtained in the heteroscedastic bimodal case suggests the
presence of a main cluster of 76 galaxies (KMM1), with the presence of
a high--velocity clump of 19 galaxies (KMM2).  In turn, the main
cluster can be subdivided in two clumps of 19 and 57 galaxies
according to the heteroscedastic trimodal case (KMM1a and KMM1b,
respectively). Table~\ref{tab1} lists the kinematical properties of
these clumps:  {\bf the three corresponding Gaussians are displayed in
Fig.~\ref{figstrip}.} Fig.~\ref{figkmm} shows that the galaxies of the KMM1a
group mainly populate the South--West central region of the cluster.
This spatial segregation suggests to investigate the velocity field in
more detail.

\input{tab1.tex}

\subsection{Velocity field}
  
The cluster velocity field may be influenced by the presence of
internal substructures, possible cluster rotation, and the presence of
other structures on larger scales, such as nearby clusters,
surrounding superclusters, and filaments. Each asymmetry effect
could produce a velocity gradient in the cluster velocity field.

\begin{figure}
\centering 
\resizebox{\hsize}{!}{\includegraphics{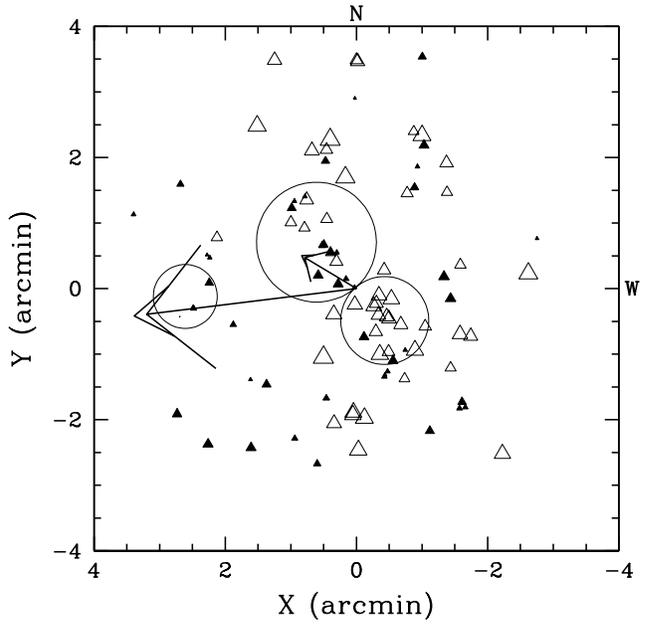}}
\caption
{Spatial distribution on the sky of the 95 member galaxies: the larger
the triangle, the smaller is the radial velocity. Open and solid triangles
indicate galaxies with velocity lower and higher than the median
cluster velocity, respectively.  The plot is centered on the cluster
center.  The big and the small arrows indicate the position angle of the
cluster gradient as measured over the whole cluster and in internal
regions, respectively.
The three circles correspond to the regions of extended
X--ray emission.}
\label{figgrad}
\end{figure}

To investigate the velocity field of \cluster we divide galaxies in a
low-- and a high--velocity samples by using the median value of galaxy
velocities $\bar{V}=250626$ \ks and check the difference between the spatial
distributions of the two samples.  High-- and low--velocity galaxies
appear segregated in the E--W direction (see Fig.~\ref{figgrad}). The
corresponding spatial distributions are different
at the $99.2\%$ c.l.  according to the two--dimensional
Kolmogorov--Smirnov test (hereafter 2DKS--test; see Fasano \&
Franceschini \cite{fas87}, as implemented by Press et
al. \cite{pre92}).

To estimate the direction of the velocity gradient we perform a
multiple linear regression fit to the observed velocities with respect
to the galaxy positions in the plane of the sky (see also den Hartog
\& Katgert \cite{den96}; Girardi et al. \cite{gir96}). We find a
position angle on the celestial sphere of $PA=97\degree\pm16\degree$ 
(measured counter--clock--wise from North), i.e. higher--velocity
galaxies lie in the East--South-East region of the cluster, in
agreement with the visual impression of galaxy distribution in
Fig.~\ref{figgrad}. To assess the significance of this velocity
gradient we perform 1000 Monte Carlo simulations by randomly shuffling
the galaxy velocities and for each simulation we determine the
coefficient of multiple determination ($RC^2$, see e.g., NAG Fortran
Workstation Handbook \cite{nag86}).  We define the significance of the
velocity gradient as the fraction of times in which the $RC^2$ of the
simulated data is smaller than the observed $RC^2$.  We find that the
velocity gradient is significant at the $98.3\%$ c.l..

We also analyze the central cluster region using 22 galaxies within a
radius of 0.4 \hh.  This choice allows us to include the position of
both X--ray peaks and exclude the East region populated by
higher--velocity galaxies only.  We find a very significant ($99.5\%$)
position angle of $PA=59\degree^{+28\degree}_{-25\degree}$,
i.e. higher--velocity galaxies lie in the direction of the North-East
X--ray clump.

\subsection{3D substructure and detection of subclumps}

The existence of correlations between positions and velocities of
cluster galaxies is a footprint of real substructures.  Here we
combine velocity and position information to compute the
$\Delta$--statistics devised by Dressler \& Schectman (\cite{dre88}).
This test is sensitive to spatially compact subsystems that 
have either an average velocity that differs from the cluster mean, or a
velocity dispersion that differs from the global one, or both.  We
find $\Delta=154$ for the value of the parameter which gives the
cumulative deviation of the local kinematical parameters (velocity and
velocity dispersion) from the global cluster parameters.  To compute
the significance of substructure we run 1000 Monte Carlo simulations,
randomly shuffling the galaxy velocities, and obtain a value of
$>99.9\%$.

\begin{figure}
\centering 
\resizebox{\hsize}{!}{\includegraphics{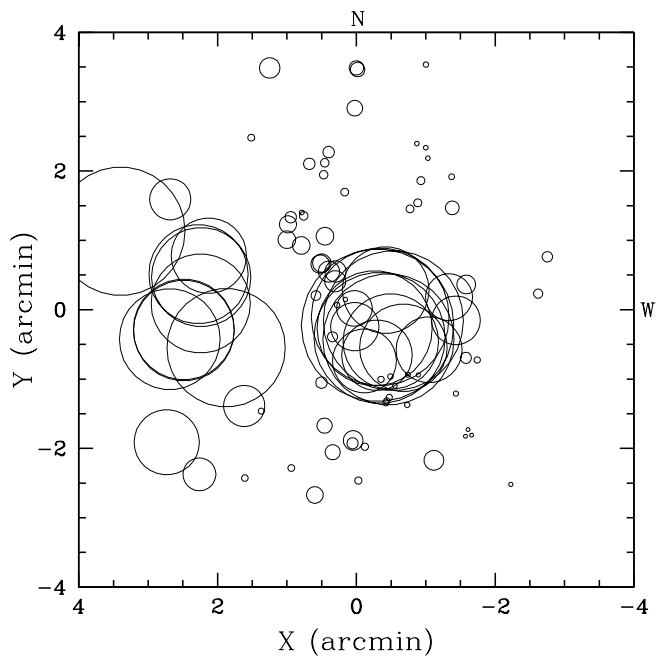}}
\caption
{Spatial distribution of cluster members, each marked by a circle: the
larger the circle, the larger is the deviation $\delta_i$ of the local
parameters from the global cluster parameters, i.e. there is more
evidence for substructure (according to the Dressler \& Schectman
test, see text).  The plot is centered on the cluster center.}
\label{figds}
\end{figure}

This technique also provides information on the positions of substructures.
Fig.~\ref{figds} shows the distribution on the sky of all galaxies,
each marked by a circle: the larger the circle, the larger the
deviation $\delta_i$ of the local parameters from the global cluster
parameters, i.e. the higher the evidence for substructure.  A clump of
galaxies with low velocity is the likely cause of large values of
$\delta_i$ in the region which lies closely at South--West of the
cluster center, i.e. in correspondence of the South-West X--ray peak.
The other possible substructure, populated by high--velocity galaxies,
lies in the Eastern region.

\begin{figure}
\centering 
\resizebox{\hsize}{!}{\includegraphics{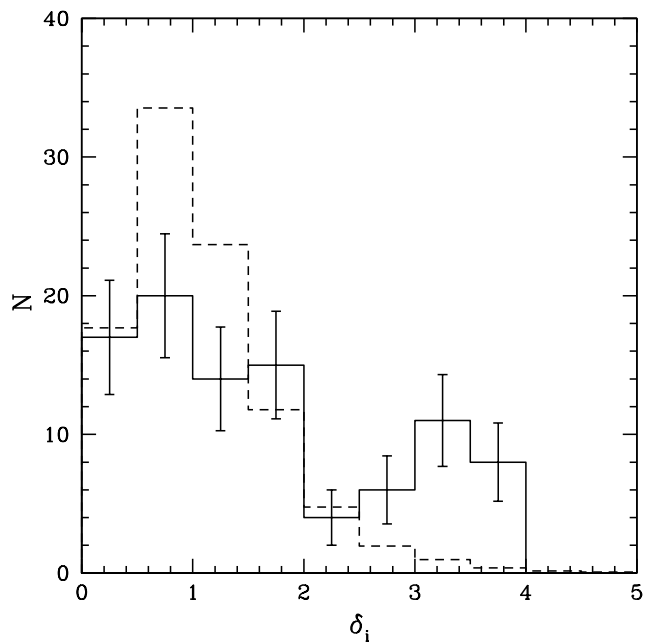}}
\caption
{The distribution of $\delta_i$ deviations of the Dressler--Schectman
analysis  for the 95
member galaxies. The solid line represents the observations,
the dashed line the distribution for the galaxies of simulated clusters,
normalized to the observed number.}
\label{figdeltai}
\end{figure}

To assign galaxies to the 3D--subclumps, we resort to the technique
developed by Biviano et al. (\cite{biv02}), who used the individual
$\delta_i$--values of the Dressler \& Schectman method. The critical
point is to determine the value of $\delta_i$ that optimally separates
between internal and external substructures. To this aim we consider the
$\delta_i$--values of all 1000 Monte Carlo simulations already used to
determine the significance of the substructure (see above).  The
resulting distribution of $\delta_i$ is compared to the observed one
finding a difference of $99.8\%$ c.l. according to the KS--test.  The
``simulated'' distribution is normalized to produce the observed
number of galaxies and compared to the observed distribution in
Fig.~\ref{figdeltai}: the latter shows a tail at large values. This
tail is populated by galaxies that presumably are in substructures.

\begin{figure}
\centering 
\resizebox{\hsize}{!}{\includegraphics{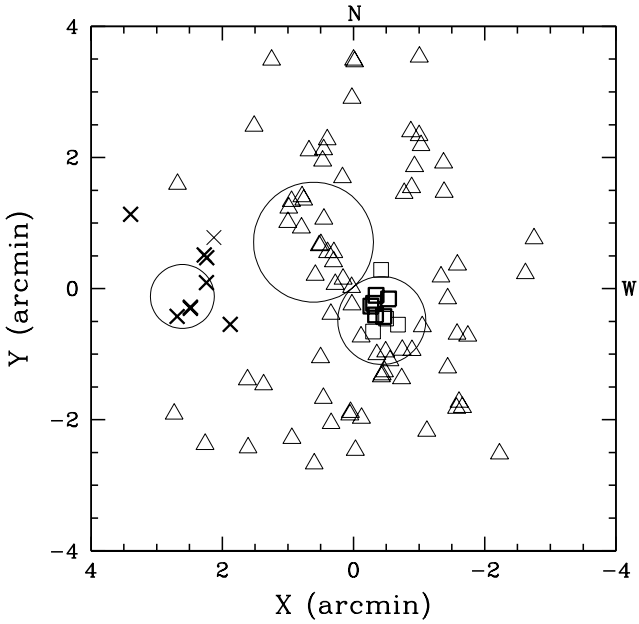}}
\caption
{Spatial distribution on the sky of the 95 member galaxies.
Squares and crosses
indicate galaxies assigned to the South--West and East clumps
detected by the Dressler--Schectman analysis,
respectively (DS-SW and DS-E): thick symbols indicate DS-SW* and DS-E*.
Triangles indicate the remaining galaxies of the main system (DS-M).}
\label{figdsi}
\end{figure}

For the selection of galaxies within substructures we choose the value
of $\delta *=3.35$, since only after the rejection of the values
$\delta_i>\delta *$, the observed and simulated distributions are no
longer distinguishable according to the KS--test.  With this choice,
14 galaxies of the cluster are assigned to substructures: six to the
central South--West clump (DS-SW*) and eight to the East clump
(DS-E*), see Fig.~\ref{figdsi}.  The velocity dispersions computed for
these structures, $\sigma_{\rm V}\simeq 300$ \ks and $\simeq 650$ \ks for
DS-S* and DS-E* clumps, respectively, are likely to be considered as
lower limits since our analysis does not guarantee the detection
of all substructure members.

We consider also a more relaxed criteria, by selecting galaxies with
$\delta_i>3$ as suggested by the histogram of Fig.~\ref{figdeltai}:
Table~\ref{tab1} shows that the results for the South--West clump
(DS-SW vs. DS-SW*) are very robust, while only one additional galaxy
in the Eastern clump (DS-E vs. DS-E*) leads to an increase of 200 \ks
in the velocity dispersion. We also consider the remaining 76 galaxies
of the main structure (DS-M). DS-M does not contain significant
structure according to the Dressler--Schectman test. However, since we
cannot exclude a residual contamination from substructure members,
its value of velocity dispersion $\sigma_{\rm V}\sim1300$ \ks is
likely to be an upper limit.

The Dressler-Schectman results superseed those of the KMM test. Again, there
is the presence of a low--velocity clump and now its South--West
position is better defined by the detection of the DS-SW clump. The
presence of a high--velocity clump is confirmed and located at the
East by the detection of the DS-E clump.  Moreover, the location of DS
clumps well coincide with X--ray peaks of extended emissions.

\subsection{Analysis of X--ray --- centered clumps}

\begin{figure}
\centering 
\resizebox{\hsize}{!}{\includegraphics{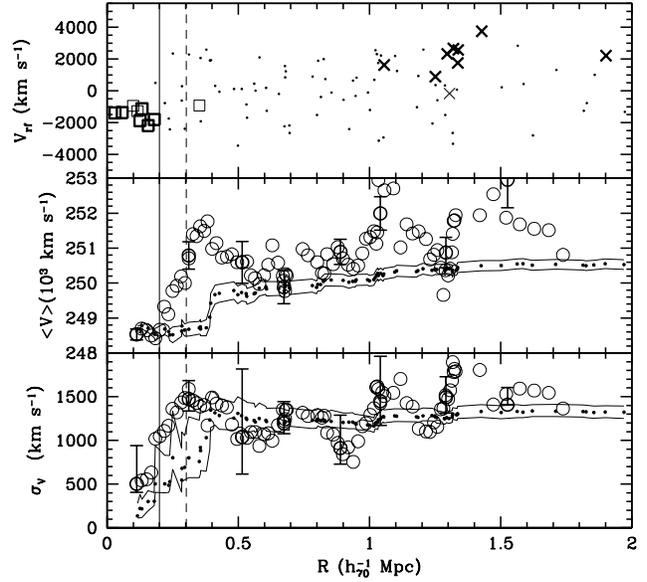}}
\caption
{ Kinematical profiles of the SW clump obtained assuming the X--ray
peak as center.  The vertical line
indicates the region likely not contaminated from other clumps (see
Sect.~3.4).  The dashed vertical line indicates the radius of the
extended X--ray emission, as defined by Demarco et al. (\cite{dem05}).
Top panel: rest--frame velocity vs. projected distance from the
clump center: squares and crosses indicate the DS-SW and DS-E as in
Fig.~\ref{figdsi}.  Differential (big circles) and integral (small
points) mean velocity and LOS velocity--dispersion profiles are shown
in middle and bottom panels, respectively. For the differential
profiles are plotted: a) the values for eight annuli from the
center of the clumps, each of 0.2 \h (heavy symbols); the current values of
each ten galaxies (faint symbols).  For the integral profiles, the
mean and dispersion at a given (projected) radius from the
clump--center is estimated by considering all galaxies within that
radius. The error bands at the $68\%$ c.l. are also shown.  }
\label{figprofxs}
\end{figure}

\begin{figure}
\centering 
\resizebox{\hsize}{!}{\includegraphics{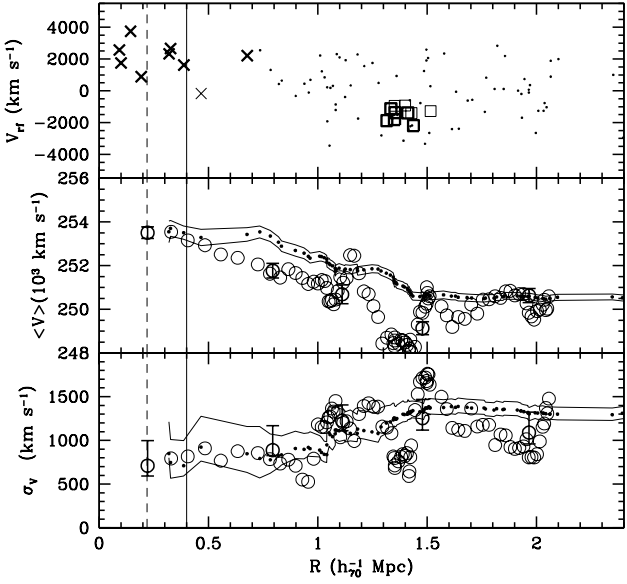}}
\caption
{ The same as in Fig.~\ref{figprofxs}, 
but {\bf referring} to the
E-clump. Note: here the annuli for the differential
profiles are 0.45 \h each.}

\label{figprofxe}
\end{figure}

\begin{figure}
\centering 
\resizebox{\hsize}{!}{\includegraphics{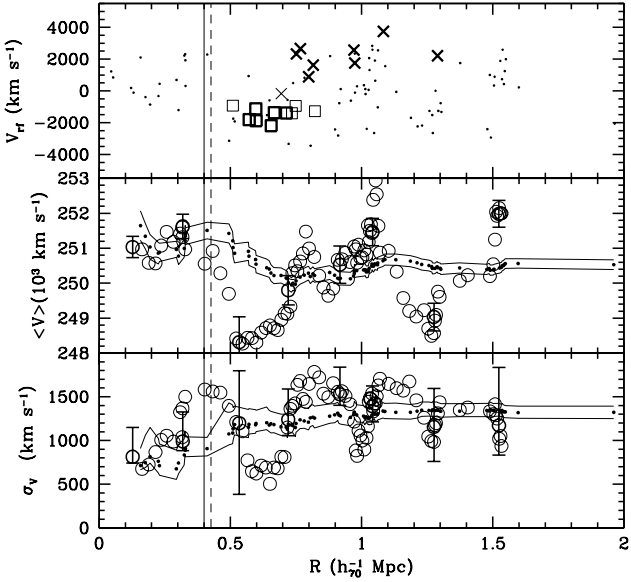}}
\caption
{ The same as in Fig.~\ref{figprofxs}, but {\bf referring} to the
central NE-clump.}
\label{figprofxn}
\end{figure}

The good spatial agreement between detected galaxy clumps and peaks of
X--ray emission prompts us to analyze the profiles of mean velocity
and velocity dispersion of galaxy systems corresponding to the
South-West, East, and North-East X--ray peaks, i.e.  using the
position of the X--ray peaks as system-centers (see
Figs.~\ref{figprofxs}, Fig.~\ref{figprofxe}, and Fig.~\ref{figprofxn},
respectively).  This allows an independent analysis of the individual
galaxy clumps.  An increase of the velocity--dispersion profile in their
central regions might be due to dynamical friction and galaxy merging
(e.g., Menci \& Fusco-Femiano \cite{men96}; Girardi et
al. \cite{gir98}; Biviano \& Katgert \cite{biv04}), or simply induced
by the presence of interlopers or of a secondary clump (e.g., Girardi
et al. \cite{gir96}). The latter hypothesis can be investigated by
looking at the
behaviour of the mean velocity profile. Figs.~\ref{figprofxs},
\ref{figprofxe}, and \ref{figprofxn} show velocity--dispersion and
mean--velocity profiles, and regions likely not contaminated by other
clumps and thus reliable for kinematical analysis. Detailed results of
this analysis are included in Table~\ref{tab1} where the clumps are
named as SW, E, and NE.

The analysis  of the South--West central region has indicated the
presence of a low--velocity clump with a low velocity--dispersion (of
300--400 \ks according to DS-SW and KMM1a results).
Fig.~\ref{figprofxs} shows how the velocity--dispersion increases with
the distance from the South-West X--ray peak.  The
mean--velocity shows a sharp change very close to the X--ray
peak, at $\sim\!0.2$ \hh. This suggests a strong contamination of
galaxies from other structures. We consider two possible 
uncontaminated regions: one within 0.2 \hh, where we find
$\sigma_{\rm V}\simeq 500$ \kss, and one within 0.18 \hh, where we
find $\sigma_{\rm V}\simeq 300$ \kss. Such a sharp change
of $\sigma_{\rm V}$ is induced just by the rejection of two galaxies,
one of which has an anomalously high velocity.  The value of
$\sigma_{\rm V}$ for the SW-clump is further analyzed in Sect.~3.5 and
discussed in Sect.~4.1.

The Dressler--Schectman analysis of the East region has indicated the
presence of a high--velocity clump with a velocity dispersion of
about 600--800 \kss. By choosing the X--ray peak as center
(Fig.~\ref{figprofxe}), the mean velocity changes at $\sim\!
0.4-0.5$ \h from the X--ray peak. Inside this region, we
obtain $\sigma_{\rm V}\sim700$ \ks for the E-clump

Fig.~\ref{figprofxn} refers to the region around the North-East X--ray
peak. The main mass clump is located in this same position, according
to the gravitational lensing analysis (Jee et al. \cite{jee05}).  We
have shown that this region is mostly populated by galaxies having
velocities intermediate between those of the above clumps, and likely
forms a high velocity dispersion structure (i.e., KMM1b clump in
Sect.~3.1, and DS-M system in Sect.~3.3).  Fig.~\ref{figprofxn} shows
an increase of the integral velocity--dispersion profile at about 0.4
\h from the X--ray peak, and a corresponding sharp change in the mean
velocity.  Moreover, galaxies of both DS-SW and DS-E substructures lie
beyond 0.4 \h from the North--East X--ray peak.  Thus, the value
$\sigma_{\rm V}\simeq 900$ \ks, computed within 0.4 \h, should be
reliable.

The three X--ray clumps differ from each other in mean velocities at a
c.l. $>99\%$, according to the means--test (e.g., Press et
al. \cite{pre92}).

Assuming that each of the three galaxy clumps is a system in dynamical
equilibrium, for each clump we compute the virial radius and the mass
contained inside with the same procedure adopted in Sect.~2 (see
Table~\ref{tab2}).  The large uncertainties associated to the mass
values are due to poor number statistics.

\input{tab2.tex}

\subsection{Spectral--type segregation}

We check for possible spectral--type segregation of galaxies, both in
position and in velocity space, by using the classification of
Demarco et al. (\cite{dem05}, see their Table~4), i.e. passive
galaxies (k), galaxies with significant Balmer lines -- likely
post-starbursts (k+a/a+k) and galaxies with relevant emission lines
 (e/k+a+[OII]).  The sample of
cluster members contains 56, 7, and 32 passive, post--starburst, and
emission--line galaxies, respectively.

\begin{figure}
\centering 
\resizebox{\hsize}{!}{\includegraphics{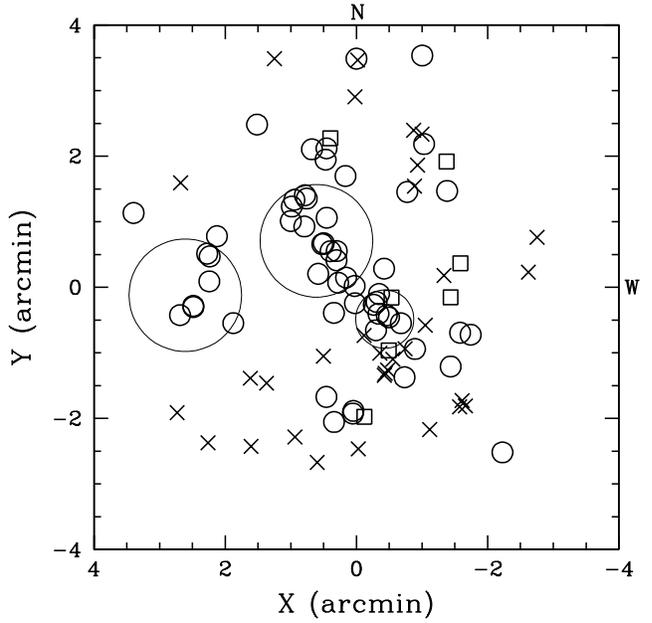}}
\caption
{Projected distribution of the 95 member galaxies.  Circles,
squares, and crosses indicate passive, post--starburst, and
emission--lines galaxies, respectively. 
 The three clumps analyzed in Sect.~3.4 are
indicated by the three circles and correspond to the regions likely not
contaminated by galaxies of other clumps,   with radii
corresponding to vertical lines in Figs.~10, 11, and 12.}
\label{figsegre}
\end{figure}

Figure~\ref{figsegre} shows the spatial distribution of galaxies of
different types. As already noted by Demarco et al., emission--line
galaxies avoid the regions of the subclumps (see also Homeier et
al. \cite{hom05}).  The same behaviour is shown by post--starburst
galaxies.  When comparing spatial distributions of passive (k) and
"active" (k+a/a+k/e/k+a+[OII]) galaxies we find a very strong
difference: $>99.99\%$, according to the 2DKS-test.

As for the velocity distributions, no difference is found between
passive and "active" galaxies, according to the KS--tests. Moreover,
mean velocities and velocity dispersions of the two populations (see
Table~\ref{tab1}) do not significantly differ according to the means--
and F--test (e.g., Press et al. \cite{pre92}).  This suggests that our
sample of member galaxies is not significantly contaminated by
interlopers. In fact, possible field galaxies would 
preferably contaminate the
sample of "active" galaxies causing a difference in the kinematical
properties with respect to the sample of passive galaxies, e.g.,
enhancing the velocity dispersion or changing the mean velocity.

\begin{figure}
\centering 
\resizebox{\hsize}{!}{\includegraphics{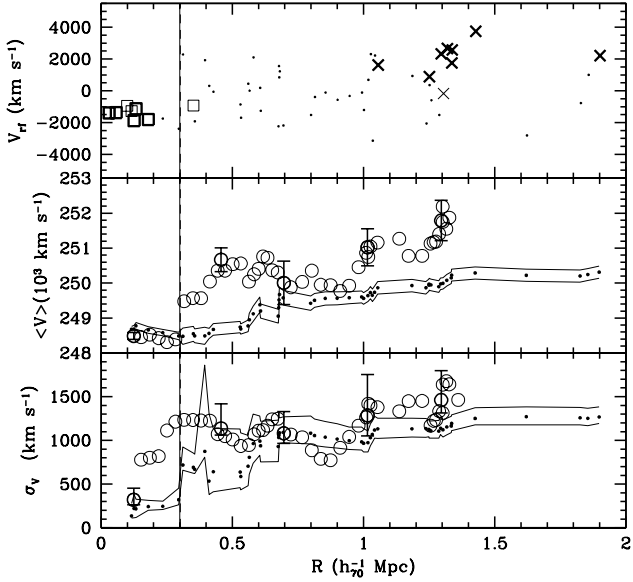}}
\caption
{ The same as in Fig.~\ref{figprofxs}, 
but considering passive galaxies only.
Here the annuli for the differential
profiles are 0.3 \hh each.}
\label{figprofxsk}
\end{figure}

Finally, we perform again the analysis of mean velocity and
velocity--dispersion profiles of Sect.~3.4, by considering passive
galaxies only. We draw different conclusions only for the
SW-clump. Fig.~\ref{figprofxsk} shows that the mean velocity now
changes only at $\sim0.3$ \h from the South-West X--ray peak and the
velocity dispersion does not increase any longer in the central
region. Within $\sim0.3$ \hh, we compute for the SW-clump a value of
$321_{-59}^{+132}$ \kss, in good agreement with the lower estimate of
$\sigma_{\rm V}$ obtained in Sect.~3.4 (see Table~\ref{tab1}).

\section{Discussion}

Out of 187 galaxies with available redshift we assign 95 members to
\clusterr. This galaxy selection is more restrictive than made by
Demarco et al. (\cite{dem05}: 102 members giving a velocity dispersion
of $\sim 1600$ \kss) due to our analysis of position and velocity
combined together. In particular, we reject a small group of galaxies
at $z=0.864$--0.867. In spite of this more restrictive member
selection, the value we obtain for the LOS velocity dispersion is
still rather high, $\sigma_{\rm V}=1322_{-68}^{+74}$ \kss, and lies in the
high--tail of the $\sigma_{\rm V}$--distribution of nearby/medium redshift
clusters (see, e.g., Fadda et al. \cite{fad96}; Mazure et
al. \cite{maz96}; Girardi \& Mezzetti \cite{gir01}). The position on 
the $L_{\rm X,bol}$--$\sigma_{\rm V}$ plane is consistent with the 
relation provided by Borgani et al. (1999) for moderately distant
clusters and by Wu et
al. (1999) for local clusters.  As for the $\sigma_{\rm V}$--$T_X$ relation, assuming the
density--energy equipartition between gas and galaxies,
i.e. $\beta_{\rm spec}=1$ (e.g., Girardi et al. \cite{gir96},
\cite{gir98}; Xue \& Wu \cite{xue00}), where
$\beta_{\rm spec}=\sigma_{\rm V}^2/(kT/\mu m_p)$ with $\mu=0.58$ the mean
molecular weight and $m_p$ the proton mass, our value of $\sigma_{\rm V}$
corresponds to $kT=10.6_{-1.1}^{+1.2}$ keV.  This value is more than
2$\sigma$ higher than the X--ray temperature determined from BeppoSAX
observations (Della Ceca et al. \cite{del00}) and more than 3$\sigma$
higher than those of the North-East and South-West X--ray systems as
determined from Chandra data (Maughan et al.  \cite{mau03}; Huo et al.
\cite{huo04}). This
suggests a strong departure from dynamical equilibrium and, in fact,
we find evidence for non--Gaussianity of the velocity distribution,
presence of a velocity gradient and significant substructure.

We find no kinematical difference between passive and "active" galaxy
populations. This suggests that our sample of member galaxies is
not significantly contaminated by interlopers. In fact, possible field
galaxies would preferably contaminate the sample of "active" galaxies causing a
difference in the kinematical properties with respect to the sample of
passive galaxies, e.g., enhancing the velocity dispersion or changing
the mean velocity.  

Instead, our analysis shows that the high value of $\sigma_{\rm V}$ is
due to the complex structure of this system, i.e. to the presence of
three galaxy clumps of different mean--velocity.  Using optical data
only we detect the low--velocity SW-clump in the central regions and
the high--velocity E-clump, which lie close to the South-West and East
peaks detected by the X--ray analysis. The North-East X--ray peak is
then associated to the main galaxy structure.  In particular, the high
relative velocity between the NE- and SW-clumps, $V_{\rm r}=(V_{\rm
NE}-V_{\rm SW})/(1+\left<z\right>)=1531$ \kss, explains the high value of
$\sigma_{\rm V}$ measured in the central cluster region and the
presence of a velocity gradient there (see Figs.~\ref{figprof} and
\ref{figgrad}), while the global velocity gradient is induced by the
presence of the high--velocity E-clump in external cluster regions.
The presence of the three galaxy clumps was already suggested by
Demarco et al. (\cite{dem05}) from the inspection of the velocity
distribution in relation to the spatial location of
galaxies. Moreover, the NE-, SW-, and E- clumps correspond to three
clumps in the mass distribution as obtained from the weak lensing analysis
(Jee et al. \cite{jee05}: C, F, and A subclumps, respectively).

As for the mass of the whole cluster, from the global analysis
of Sect.~2 we obtain $M(<2.0 \hhh)=(2.2\pm0.6)\times 10^{15}$\mm.
Since the system is not virialized, but likely bound (see the
discussion below), this estimate might overestimate the mass even
by a factor two.  Adding the mass estimates of each clump within its
virial radius (see Table~\ref{tab2}, Sect.~3.4), we obtain
$M=1.2_{-0.3}^{+0.5}\times 10^{15}$\mm: this estimate should be
considered as a lower value within 2.0 \h, since
it does not consider other small clumps or isolated infalling, bound
galaxies, as well the likely possibility that the three clumps extend
outside the virial radius.  Thus, we conclude that the cluster mass
within 2 Mpc lies in the range of $(1.2-2.2)\times 10^{15}$\mm, which
is typical to that of very massive clusters (e.g., Girardi et al. \cite{gir98};
Girardi \& Mezzetti \cite{gir01}).

Our mass estimate is consistent with that of Maughan et
al. (\cite{mau03}) of $(1.1\pm0.2)\times 10^{15}$\mm, based on Chandra
X--ray analysis and considering only the two central clumps within 1.4
\hh.  To compare with results from weak lensing analyses we also
compute the projected mass, by considering the global cluster geometry
as formed by the three clumps at the cluster redshift. Each clump is
described by the King--like mass distribution (see Sect.~2) or,
alternatively, by a NFW profile where the mass--dependent
concentration parameter is taken from Navarro et al. (\cite{nav97})
and rescaled by the factor 1+z (Bullock et al. \cite{bul01}; Dolag et
al. \cite{dol04}). The mass distribution of each clump is truncated at
one virial radius or, alternatively, at two virial radii.  In the
following we indicate the range of our results.  We find the projected
mass within 1 Mpc from the center of the main clump (NE-clump) to be
$(9-15)$\mquaa, higher than that, 5\mqua, of Jee et
al. (\cite{jee05}), but in agreement with the value $\gtrsim 1\times
10^{15}$\m of Huo et al. (\cite{huo04}, see their Figure 10).  Indeed,
both Huo et al.  and Jee et al. compare their weak lensing results
with an isothermal sphere with $\sigma_{\rm V}=$900--1000 \kss, in
agreement with the value of $\sigma_{\rm V}$ that we measure for the
main galaxy clump. However, the weak--lensing mass lies above or below
the isothermal sphere mass for Jee et al.  and Huo et al.,
respectively.

\subsection{Individual clumps}

Our estimate of $\sigma_{\rm V}$ for the NE-clump well agrees with
that of Demarco et al. (\cite{dem05}) and corresponds to
$kT=4.8_{-0.4}^{+1.8}$ keV, in agreement with the observed gas
temperature of $\sim 6$ KeV (Maughan et al.  \cite{mau03}; Tozzi et
al. \cite{toz03}; Huo et al. \cite{huo04}).  Similarly, our mass
estimate, $M(<R_{\rm vir}=1.35 \hhh) =7.0_{-2.1}^{+2.9}$\mqua, well
agrees with the X--ray mass by Maughan et al. [\cite{mau03}, $M(<1.4
\hhh)=7.0_{-1.5}^{+1.7}$\mquaa].  To compare our results with other
studies we rescale $M(<R_{\rm vir})$ at their radii by using the
King--like profile or, alternatively, the NFW profile (see above).  In
the following we give the two values obtained from the rescaling,
reliable with a $30\%$ lower--error and a $40\%$ upper--error, as
derived from the estimate of $M(<R_{\rm vir})$. Our estimates well agree with
those of other studies: $M(<0.43 \hhh)=$(1.9--2.6)\mquaa, cf. with
$(2.5\pm 0.9)$\mqua by Demarco et al. (\cite{dem05}), based on galaxy
dynamics; $M(r<65\dotarcsss=0.496 \hhh)=$(2.3--2.9)\mquaa, cf. with
$(3\pm1)$\mqua by Joy et al. (\cite{joy01}), based on the
Sunyaev--Zeldovich effect; $M(r<0.753 \hhh)$=(3.8--4.3)\mquaa,
cf. with $(2.66\pm 0.77)$\mqua by Ettori et al. (\cite{ett04}), based
on Chandra X--ray data; $M(r<1 \hhh)=$(5.1--5.4)\mquaa, cf. with
$\sim 5$\mqua by Huo et al. (\cite{huo04}), based on Chandra X--ray
data.

As for the SW-clump, the results in the literature are not yet
clear. In fact, the X--ray temperature suggests that the North-East
and the South-West clumps are similar in mass (Maughan et
al. \cite{mau03}; Huo et al. \cite{huo04}), while both the optical
analysis by Demarco et al. (\cite{dem05}) and the weak lensing
analysis by Jee et al. (\cite{jee05}) find that the South-West clump
is about half massive than the North-East clump.  Our analysis of the
$\sigma_{\rm V}$--profile gives two alternative values for
$\sigma_{\rm V}$: the larger value is consistent with that found 
by Demarco et al. (\cite{dem05}, cf. $\sigma_{\rm V}=503_{-96}^{+439}$
\ks vs.  their $737\pm126$ \kss) and with the observed gas
temperatures of 5-6 keV (Maughan et al. \cite{mau03}; Huo et
al. \cite{huo04}), while the lower estimate, $\sigma_{\rm
V}=301_{-107}^{+122}$, is significantly different. This uncertainty
is due to the fact that the $\sigma_{\rm V}$ profile increases in
central regions (see Fig.\ref{figprofxs}) and thus the $\sigma_{\rm
V}$ estimate strongly depends on the considered region.  Demarco et
al.  considered a region (based on X--ray data) larger than our region
(based on kinematical data). Our analysis of passive galaxies also
gives a small value, $\sigma_{\rm V}\sim 300$ \ks, thus suggesting two
alternative hypothesis: 1) high values of $\sigma_{\rm V}$ are due to
galaxy--contamination by other clumps, so that the SW-clump should be
considered as a very small group, 2) we are dealing with
 a very relaxed core hosted in a high--$\sigma_{\rm V}$,
massive cluster.  The second hypothesis is consistent with the
observations of nearby clusters where $\sigma_{\rm V}$ of the
subsample of bright central elliptical galaxies is lower than
$\sigma_{\rm V}$ of the whole cluster (Biviano \& Katgert
\cite{biv04}), a phenomenon possibly due to dynamical friction and
galaxy merging (e.g., Menci \& Fusco-Femiano \cite{men96}).  Only a
deeper galaxy sample would allow us to better trace and separate the
North--East and the South-West systems and thus discriminate between
the two hypotheses. However, the SW-clump appears to be so dense of
galaxies that we are inclined to believe in the detection of a
cluster--core.  In this case,  we note
that: a) our mass estimate would be an underestimate of the global
mass of the Southern cluster; b) our results would be reconciled with
high values of gas temperature and X--ray luminosity (Maughan et
al. \cite{mau03}; Tozzi et al. \cite{toz03}; Huo et al. \cite{huo04}).

As for the Eastern clump, the level of X--ray emission in the Chandra
image is much lower than those of the North-East or the South-West
clumps (see Fig.~1 of Demarco et al. \cite{dem05}). On the contrary,
its gravitational--lensing mass is comparable to that of the
South-West clump (see A and F clumps by Jee et al. \cite{jee05}), and
our estimate of velocity dispersion is typical of a massive cluster,
$\sigma_{\rm V}\sim 700$ \kss.  This discrepancy with X--ray
luminosity suggests that this {\bf galaxy system} is far from being
virialized, maybe elongated along the LOS (thus giving a high
$\sigma_{\rm V}$ and a high projected lensing mass), with the gas
component not very dense.  {\bf In particular, the Eastern X--ray peak
might be associated to a small group embedded in a large--scale
structure filament connecting to the cluster from the East--South-East
region, which is populated by higher velocity --maybe more distant --
galaxies (see Fig.~\ref{figgrad}).}  In the case of a bound, but non
virialized structure, we might have overestimated the mass even by a
factor two.

We finally compare the projected mass of the three clumps within a
radius of $20\dotarcsss$ $\ $ with the results from weak
lensing by Jee et al. (\cite{jee05}).  The resulting values for
projected masses of the NE-, SW-, and E-clumps lie in the ranges
$(1.6-2.3)$\mquaa, $(0.5-0.7)$\mquaa, and $(1.0-1.5)$\mquaa, all
values somewhat higher than those reported in Table2 of Jee et al. for
clumps C, F, and A, respectively.

\subsection{Analytic calculations of the  dynamical status}

Here, we investigate the relative dynamics of the NE- and SW-clumps in
the central cluster region using different analytic approaches which
are based on an energy integral formalism in the framework of locally
flat spacetime and Newtonian gravity (e.g., Beers et
al. \cite{bee82}).  The values of the relevant observable quantities
for the two--clump system are: the relative LOS velocity, $V_{\rm
r}=1531$ \kss, the projected linear distance between the two clumps,
$D=0.66$ \hh, the mass of the system obtained adding the masses of the
two clumps each within its virial radius, $M_{\rm
sys}=8.2_{-2.2}^{+3.6}$\mquaa.

\begin{figure}
\centering 
\resizebox{\hsize}{!}{\includegraphics{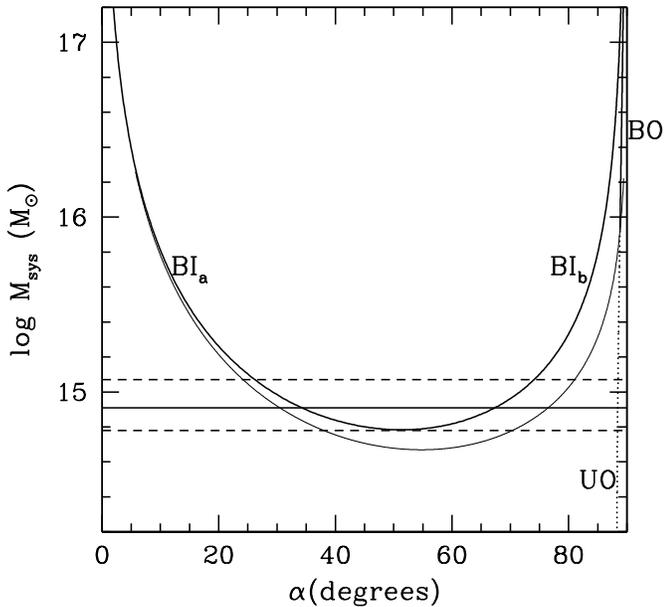}}
\caption[...]{System mass vs. projection angle for bound and unbound
solutions of the two--body model applied to the NE- and SW-clumps (solid
and dotted curves, respectively, see text).  The thin curve separates
the bound and unbound regions according to the Newtonian criterion
(above and below the curve, respectively).  The horizontal lines
give the observational values of the mass system and its 1$\sigma$
error bands.}
\label{figbim}
\end{figure}

First, we consider the Newtonian criterion for gravitational binding
stated in terms of the observables as $V_{\rm r}^2D\leq2GM_{\rm
sys}sin^2\alpha cos\alpha$, where $\alpha$ is the projection angle
between the plane of the sky and the line connecting the centers of
two clumps.  The faint curve in Fig.~\ref{figbim} separates the bound
and unbound regions according to the Newtonian criterion (above and
below the curve, respectively). Considering the value of $M_{\rm sys}$,
the NE+SW system is bound between $30\degree$ and $77\degree$: the
corresponding probability, computed considering the solid angles (i.e.,
$\int^{77}_{30} cos\,\alpha\,d\alpha$), is $47\%$.  We also consider
the implemented criterion $V_{\rm r}^2D\leq2GMsin^2\alpha_{\rm V}
cos\alpha_D$, which introduces different angles for projection of
distance and velocity, not assuming strictly radial motion between the
clumps (Hughes et al. \cite{hug95}).  We obtain a binding probability
of $44\%$.

Then, we apply the analytical two--body model introduced by Beers et
al. (\cite{bee82}) and Thompson (\cite{tho82}; see also Lubin et
al. \cite{lub98} for a recent application).  This model assumes radial
orbits for the clumps, with no shear or net rotation of the
system. Furthermore, the clumps are assumed to start their evolution
at time $t_0=0$ with separation $d_0=0$, and are moving apart or
coming together for the first time in their history, i.e. we are
assuming that we are seeing the cluster prior to merging.  The bimodal
model solution gives the total system mass $M_{\rm sys}$ as a function
of $\alpha$ (e.g., Gregory \& Thompson \cite{gre84}).
Fig.~\ref{figbim} compares the bimodal--model solutions with the
observed mass of the system, which is the most uncertain observational
parameter.  The present bound outgoing solutions (i.e. expanding), BO,
are clearly inconsistent with the observed mass.  The possible
solutions span these cases: the bound and present incoming solution
(i.e. collapsing), $BI_a$ and $BI_b$, and the unbound-outgoing
solution, $UO$.  For the incoming case there are two solutions because
of the ambiguity in the projection angle $\alpha$.  We compute the
probabilities associated to each solution assuming that the region of
$M_{\rm sys}$ values between 1$\sigma$ bands is equally probable for
individual solutions: $P_{\rm BIa}\sim65\%$, $P_{\rm BIb}\sim35\%$,
$P_{\rm UO}<1\times 10^{-4}\%$.

There are several limitations to characterize the dynamics
of the central region of \cluster using these models. For instance,
possible underestimates of the masses, e.g., if the clumps extend
outside the virial radius or if the SW-clump is only the core of the
South-West system (see above), lead to binding probabilities larger
than those computed above.  Moreover, the models do not
take into accounts the mass distribution in the clumps when the
separation of the clumps is comparable with their size (i.e. at small
$\alpha$) and do not consider the possible effect of the E-clump.
Finally, the two--body model breaks down in a regime where the NE- and
SW-clumps are already strongly interacting, as suggested by several
evidences: the displacement between peaks of gas distribution and of
galaxy/dark matter distribution (Maughan et al.  \cite{mau03}; Huo et
al. \cite{huo04}); Jee et al. \cite{jee05}); the possible presence of
a shock front (Maughan et al.  \cite{mau03}); the presence of galaxies
showing a very recent star formation episode (J\o rgensen et
al. \cite{jor05});
 the segregation of star--forming and non star--forming
galaxies probably induced by the   
interaction with the intra--cluster medium (Homeier et al. \cite{hom05}).

  Looking at galaxies only we cannot discriminate
between a pre-- or post--merging phase since the galaxy component is
very robust against mergers, e.g., two clusters can pass through one
another without destroying the individual optical components (e.g.,
White \& Fabian \cite{whi95}; Roettiger et al.  \cite{roe97}). Note,
for instance, that the properties of the SW-clump resemble those of
cluster--cores destined to survive tidal disruption during the merger:
size comparable to the cluster core and mass $\lesssim 0.05\times$
cluster mass (see Gonz\'alez--Casado et al. \cite{gon94}).  These
cores will be detectable in the host cluster as a substructure for a
long time.  Since the gas component shows two well distinct entities
in the central cluster regions, we presume that the merging is not too
advanced, i.e. well before the coalescence.

Under the assumption that the two central clumps are already very
close, we apply the above dynamical models to the system made of the
[(NE+SW)+E] clumps, too. The values of the relevant observable
quantities are: $V_{\rm r}=1401$ \kss, $D=1.09$ \hh, and $M_{\rm
sys}=11.7_{-2.6}^{+4.7}$ \mquaa.  We obtain that the binding
probabilities are $48\%$, and $45\%$, according to the Newtonian
criterion and its implementation, respectively; while the two--body
model gives a probability $>99.9\%$ for the bound incoming solution.

\section{Summary \& conclusions}

We present the results of the dynamical analysis of the cluster of
galaxies \clusterr, one of the most massive structures known at
$z>0.8$.  The X--ray emission is known to have two clumps in the
central regions, and a third clump $\sim\!1$ Mpc to the East.  Our
analysis is based on velocities and positions of member galaxies taken
from the extensive spectroscopic survey performed by Demarco et
al. (\cite{dem05}), i.e.  187 galaxies having redshift in the cluster
region.

We find that \cluster appears as a well isolated peak in the redshift
space at $z=0.836$, and select 95 cluster members.  We compute a value
for the LOS velocity dispersion of galaxies,
$\sigma_{\rm V}=1322^{+74}_{-68}$ \kss, much larger than expected
for a relaxed cluster with an observed X--ray temperature of
$\sim 5-6$ keV.

We find evidence that this cluster is far from dynamical equilibrium,
as shown by:

\begin{itemize}

\item the non Gaussianity of the velocity distribution according to
different tests, at the 90--98$\%$ c.l., as well
as the presence of significant velocity gaps;

\item the correlation between velocities and positions of galaxies at the
$>99\%$ c.l., and the presence of a velocity gradient;

\item the presence of significant substructures at the $>99.9\%$ c.l..

\end{itemize}

To detect and analyze possible subsystems we used different methods.

\begin{itemize}

\item By applying the KMM method we find that a two--clumps, and likely 
a three--clumps partition of the velocity distribution
is significantly better than a single Gaussian to describe
the velocity distribution; in particular, the galaxies of KMM1a group
are mainly located in the South--West central region. 

\item By combining positions and velocities in the Dressler \&
Schectman statistics we detect two substructures, well corresponding
in location to the South--West and East X--ray peaks, in addition to
the main cluster component identified with the North--East X--ray peak.

\item 
Taking advantage of X--ray peak determination, we analyze the three
galaxy clumps centered in these peaks through the profiles of mean velocity
and velocity dispersion. This analysis allows us to estimate the clump
region that is likely not contaminated by galaxies of other clumps
and to evaluate the kinematical properties.

\end{itemize}

In summary, our analysis shows that the high value of $\sigma_{\rm V}$
is due to the complex structure of \clusterr, i.e. to the presence of
three galaxy clumps of different mean--velocity.  Using optical data
we detect a low--velocity clump (with $\sigma_{\rm V}=$300--500
\kss) in the central South--West region and a high--velocity clump
(with $\sigma_{\rm V}\simeq$700 \kss) in the Eastern region, nicely
matching the position of the South--West and East peaks detected in
the X--ray emission.  The central North--East X--ray peak is
associated to the main galaxy structure having intermediate velocity
and $\sigma_{\rm V}\sim 900$ \kss.  The three clumps differ from each
other in mean
velocities at a c.l. $>99\%$ (relative LOS velocities are
$>1000$ \kss).

The mass of the whole system within 2 \h is estimated to be
(1.2--2.2)$\times 10^{15}$\mm, where the upper and lower
limits come from the virial analysis of the cluster as a whole and
from the sum of virial masses of the three individual clumps,
respectively.

Analytic calculations, based on the two-body model, indicate that the
system is most likely bound, destined to merge.  In particular, we
suggest that the South--West clump is not a small group, but rather
the dense core of a massive cluster, able to survive tidal disruption
during the merger.

In conclusion, \cluster reveals a very complex structure, with several
clumps likely destined to merge in a very massive cluster.
Our results lend further support to the picture that massive clusters at 
$z>0.8$ are dynamically complex and, therefore, likely to be
young. This indicates that we are approaching the epoch at which such
massive structures take shapes from the evolution of the cosmic web.
On-going extensive spectroscopic surveys of such systems at $z\sim\!
1$ and beyond, combined with detailed analyses of their gaseus and 
dark matter
components (now possible with weak lensing analysis of
HST-ACS data; Jee et al. \cite{jee05}; Lombardi et al. \cite{lom05}), will
shed new light on cluster formation processes.

\begin{acknowledgements}
We thank Andrea Biviano, Massimo Ramella, and Paolo Tozzi for useful
discussions.  We are grateful to the anonymous referee for helpful
comments. This work has been partially supported by the 
Italian Space Agency (ASI), by the Istituto Nazionale di Astrofisica
(INAF) through grant D4/03/IS, and by the Istituto Nazionale di Fisica
Nucleare (INFN) through grant PD-51.

\end{acknowledgements}

\end{document}

%% file: tab1.tex
\begin{table}
        \caption[]{Results of Kinematical analysis}
         \label{tab1}
                $$
         \begin{array}{l r l l}
            \hline
            \noalign{\smallskip}
            \hline
            \noalign{\smallskip}
\mathrm{Sample} & \mathrm{N_g} & \phantom{249}\mathrm{<V>}\phantom{249} & 
\phantom{24}\sigma_V^{\mathrm{a}}\phantom{24}\\
& &\phantom{249}\mathrm{km\ s^{-1}}\phantom{249} 
&\phantom{2}\mathrm{km\ s^{-1}}\phantom{24}\\
            \hline
            \noalign{\smallskip}
 
\mathrm{Whole\ system} & 95 &250530\pm135 &1322_{-68}^{+74}\\
\hline
\mathrm{KMM\ partitions}&&\\
\hline
\mathrm{KMM1} & 76 &249730\pm124 &1080_{-53}^{+113}\\
\mathrm{KMM2} & 19 &253634\pm50  &\phantom{1}210_{-22}^{+31}\\ 
\mathrm{KMM1a}& 19 &247290\pm97  &\phantom{1}408_{-50}^{+67}\\
\mathrm{KMM1b}& 57 &250498\pm102 &\phantom{1}768_{-50}^{+97}\\
\hline
\mathrm{Dressler-Schectman\ structures}&&\\
\hline
\mathrm{DS-SW*}&  6 &248338\pm146 &\phantom{1}318_{-36}^{+96}\\
\mathrm{DS-E*} &  8 &253488\pm241 &\phantom{1}645_{-110}^{+263}\\
\mathrm{DS-SW} & 10 &248613\pm107 &\phantom{1}317_{-47}^{+81}\\
\mathrm{DS-E}  &  9 &253429\pm298 &\phantom{1}848_{-183}^{+330}\\
\mathrm{DS-M}  & 76 &250510\pm149 &1293_{-69}^{+94}\\
\hline
\mathrm{Properties\ of\ X-ray - centered\ clumps}&&\\
\hline
\mathrm{SW-clump(<0.2 Mpc)}   & 10 &248535\pm172 
&\phantom{1}503_{-96}^{+439}\\
\mathrm{SW-clump(<0.18 Mpc)} & 8  &248713\pm121 
&\phantom{1}301_{-107}^{+122}\\
\mathrm{E-clump(<0.4 Mpc)}    &  7 &253506\pm304 
&\phantom{1}710_{-117}^{+287}\\
\mathrm{NE-clump(<0.4 Mpc)}   & 15 &251346\pm241 
&\phantom{1}888_{-75}^{+152}\\
\hline
\mathrm{Segregation\ analysis}&&\\
\hline
\mathrm{passive\ galaxies}       & 56 &250313\pm171 &1268_{-81}^{+114}\\
\mathrm{"active"\ galaxies}      & 39 &250803\pm226 &1410_{-125}^{+139}\\
\mathrm{SW-clump(passive\ gals)(<0.3 Mpc)}   & 10 &248488\pm110 
&\phantom{1}321_{-59}^{+132}\\
              \noalign{\smallskip}
            \hline
            \noalign{\smallskip}
            \hline
         \end{array}
$$
\begin{list}{}{}  
\item[$^{\mathrm{a}}$] We use the biweigth and the gapper estimators by
Beers et al. (1990) for samples with $\mathrm{N_g}\ge$ 15 and with
$\mathrm{N_g}<15$ galaxies, respectively (see also Girardi et
al. \cite{gir93}).
\end{list}
         \end{table}

%% file: tab2.tex
\begin{table}
        \caption[]{Virial mass estimates}
         \label{tab2}
$$
           \begin{array}{l r l r}
            \hline
            \noalign{\smallskip}
            \noalign{\smallskip}
\mathrm{Sample} & R_{\mathrm{vir}} & M(<R_{\mathrm{vir}})\\
            \noalign{\smallskip}
& \mathrm{Mpc}  & 10^{14}\times M_{\odot}\\
            \hline
            \hline
            \noalign{\smallskip}
 
\mathrm{Whole\ system} & 2.0 &\phantom{2}22\pm6\\
\mathrm{SW-clump^a} & 0.8  &\phantom{2}1.2_{-0.5}^{+2.1}\\
\mathrm{SW-clump^b}& 0.5  &\phantom{2}0.27_{-0.22}^{+0.23}\\
\mathrm{E-clump}  & 1.1  &\phantom{2}3.5_{-1.5}^{+3.0}\\
\mathrm{NE-clump} & 1.3  &\phantom{2}7.0_{-2.1}^{+2.9}\\
              \noalign{\smallskip}
            \hline
            \noalign{\smallskip}
            \hline
         \end{array}
$$
\begin{list}{}{}
\item[$^{\mathrm{a}}$] Using the $\sigma_V$ computed within 0.2 \h
(see Table~\ref{tab1}).
\item[$^{\mathrm{b}}$] Using the $\sigma_V$ computed within 0.18 \h
(see Table~\ref{tab1}).
\end{list}
         
\end{table}